\documentclass[twocolumn,showpacs,aps,prd,superscriptaddress,floatfix]{revtex4}

\usepackage{graphicx}
\usepackage{dcolumn}
\usepackage{bm}

\RequirePackage{xspace}

\usepackage{relsize}
\def\babar{\mbox{\slshape B\kern-0.1em{\smaller A}\kern-0.1em
    B\kern-0.1em{\smaller A\kern-0.2em R}}}

\def\epem       {\ensuremath{e^+e^-}\xspace}

\def\mtau       {\ensuremath{\tau}\xspace}

\def\taup       {\ensuremath{\tau^+}\xspace}

\def\tautau     {\ensuremath{\tau^+\tau^-}\xspace}

\def\nub        {\ensuremath{\overline{\nu}}\xspace}

\def\nub        {\ensuremath{\overline{\nu}}\xspace}

\def\q     {\ensuremath{q}\xspace}
\def\qbar  {\ensuremath{\overline q}\xspace}

\def\u     {\ensuremath{u}\xspace}

\def\d     {\ensuremath{d}\xspace}

\def\s     {\ensuremath{s}\xspace}

\def\c     {\ensuremath{c}\xspace}

\def\piz   {\ensuremath{\pi^0}\xspace}

\def\Kbar  {\kern 0.2em\overline{\kern -0.2em K}{}\xspace}

\def\Kz    {\ensuremath{K^0}\xspace}
\def\Kzb   {\ensuremath{\Kbar^0}\xspace}
\def\KzKzb {\ensuremath{\Kz \kern -0.16em \Kzb}\xspace}
\def\Kp    {\ensuremath{K^+}\xspace}
\def\Km    {\ensuremath{K^-}\xspace}

\def\KpKm  {\ensuremath{\Kp \kern -0.16em \Km}\xspace}
\def\KS    {\ensuremath{K^0_{\scriptscriptstyle S}}\xspace}

\def\Dbar    {\kern 0.2em\overline{\kern -0.2em D}{}\xspace}

\def\Dz      {\ensuremath{D^0}\xspace}
\def\Dzb     {\ensuremath{\Dbar^0}\xspace}
\def\DzDzb   {\ensuremath{\Dz {\kern -0.16em \Dzb}}\xspace}
\def\Dp      {\ensuremath{D^+}\xspace}
\def\Dm      {\ensuremath{D^-}\xspace}

\def\DpDm    {\ensuremath{\Dp {\kern -0.16em \Dm}}\xspace}

\def\Dstarz  {\ensuremath{D^{*0}}\xspace}

\def\B       {\ensuremath{B}\xspace}
\def\Bbar    {\kern 0.18em\overline{\kern -0.18em B}{}\xspace}

\def\BB      {\ensuremath{B\Bbar}\xspace} 
\def\Bz      {\ensuremath{B^0}\xspace}
\def\Bzb     {\ensuremath{\Bbar^0}\xspace}
\def\BzBzb   {\ensuremath{\Bz {\kern -0.16em \Bzb}}\xspace}
\def\Bu      {\ensuremath{B^+}\xspace}
\def\Bub     {\ensuremath{B^-}\xspace}
\def\Bp      {\ensuremath{\Bu}\xspace}
\def\Bm      {\ensuremath{\Bub}\xspace}

\def\BpBm    {\ensuremath{\Bu {\kern -0.16em \Bub}}\xspace}

\def\BorBbar    {\kern 0.18em\optbar{\kern -0.18em B}{}\xspace}
\def\DorDbar    {\kern 0.18em\optbar{\kern -0.18em D}{}\xspace}
\def\KorKbar    {\kern 0.18em\optbar{\kern -0.18em K}{}\xspace}

\def\jpsi     {\ensuremath{{J\mskip -3mu/\mskip -2mu\psi\mskip 2mu}}\xspace}

\mathchardef\Upsilon="7107
\def\Y#1S{\ensuremath{\Upsilon{(#1S)}}\xspace}

\def\FourS {\Y4S}

\mathchardef\Deltares="7101
\mathchardef\Xi="7104
\mathchardef\Lambda="7103
\mathchardef\Sigma="7106
\mathchardef\Omega="710A

\def\Deltabar{\kern 0.25em\overline{\kern -0.25em \Deltares}{}\xspace}
\def\Lbar{\kern 0.2em\overline{\kern -0.2em\Lambda\kern 0.05em}\kern-0.05em{}\xspace}
\def\Sigbar{\kern 0.2em\overline{\kern -0.2em \Sigma}{}\xspace}
\def\Xibar{\kern 0.2em\overline{\kern -0.2em \Xi}{}\xspace}
\def\Obar{\kern 0.2em\overline{\kern -0.2em \Omega}{}\xspace}
\def\Nbar{\kern 0.2em\overline{\kern -0.2em N}{}\xspace}
\def\Xb{\kern 0.2em\overline{\kern -0.2em X}{}\xspace}

\def\mes        {\mbox{$m_{\rm ES}$}\xspace}

\def\DeltaE     {\mbox{$\Delta E$}\xspace}

\newcommand{\tev}{\ensuremath{\mathrm{\,Te\kern -0.1em V}}\xspace}
\newcommand{\gev}{\ensuremath{\mathrm{\,Ge\kern -0.1em V}}\xspace}
\newcommand{\mev}{\ensuremath{\mathrm{\,Me\kern -0.1em V}}\xspace}
\newcommand{\kev}{\ensuremath{\mathrm{\,ke\kern -0.1em V}}\xspace}
\newcommand{\ev}{\ensuremath{\mathrm{\,e\kern -0.1em V}}\xspace}
\newcommand{\gevc}{\ensuremath{{\mathrm{\,Ge\kern -0.1em V\!/}c}}\xspace}
\newcommand{\mevc}{\ensuremath{{\mathrm{\,Me\kern -0.1em V\!/}c}}\xspace}
\newcommand{\gevcc}{\ensuremath{{\mathrm{\,Ge\kern -0.1em V\!/}c^2}}\xspace}
\newcommand{\mevcc}{\ensuremath{{\mathrm{\,Me\kern -0.1em V\!/}c^2}}\xspace}

\def\invfb   {\ensuremath{\mbox{\,fb}^{-1}}\xspace}

\def\mus  {\ensuremath{\rm \,\mus}\xspace}

\def\mus        {\ensuremath{\,\mu{\rm s}}\xspace}    

\def\to                 {\ensuremath{\rightarrow}\xspace}

\def\pep2{PEP-II}

\def\gsim{{~\raise.15em\hbox{$>$}\kern-.85em
          \lower.35em\hbox{$\sim$}~}\xspace}
\def\lsim{{~\raise.15em\hbox{$<$}\kern-.85em
          \lower.35em\hbox{$\sim$}~}\xspace}

\def\Vub  {\ensuremath{|V_{ub}|}\xspace}

\xspace

\newcommand{\jprlBase}       {Phys.\ Rev.\ Lett.\xspace}
\newcommand{\jprBase}        {Phys.\ Rev.\xspace}
\newcommand{\jplBase}        {Phys.\ Lett.\xspace}
\newcommand{\nimBaseA}       {Nucl.\ Instr.\ Methods Phys.\ Res., Sect.\ A\xspace}

\newcommand{\npBase}         {Nucl.\ Phys.\xspace}

\newcommand{\jpg}       [1]  {{J.\ Phys.\ {\bf G{\bf #1}}}}

\newcommand{\nima}      [1]  {\nimBaseA~{\bf #1}}

\newcommand{\npb}       [1]  {\npBase\ B~{\bf #1}}

\newcommand{\plb}       [1]  {\jplBase\ B~{\bf #1}}

\newcommand{\jprl}      [1]  {\jprlBase\ {\bf #1}}
\newcommand{\jprd}      [1]  {\jprBase\ D~{\bf #1}}

\newcommand{\progtp}    [1]  {{Prog.\ Theor.\ Phys.\ {\bf #1}}}

\def\jetset74   {\mbox{\tt Jetset \hspace{-0.5em}7.\hspace{-0.2em}4}\xspace}

\def\btn {\ensuremath{B^{+} \to \tau^{+} \nu}\xspace}

\def\eextra {\ensuremath{E_{\mathrm{extra}}}\xspace}

\def\tautoenunu {\ensuremath {\tau^+ \to e^+ \nu \nub}}

\def\tautomununu {\ensuremath {\tau^+ \to \mu^+ \nu \nub}}

\def\tautopinu {\ensuremath {\tau^+ \to \pi^+ \nub}}

\def\onlumi    {\ensuremath { 426  \invfb\ }}

\newcommand{\BABARPubYear}    {12}
\newcommand{\BABARPubNumber}  {020}

\newcommand{\SLACPubNumber} {15134}

\def\geantfour      {\mbox{\tt GEANT4}\xspace}
\newcommand{\tauex}{\ensuremath{\tau^+\to e^+\nu\bar{\nu}}}
\newcommand{\taumux}{\ensuremath{\tau^+\to \mu^+\nu\bar{\nu}}}
\newcommand{\taupix}{\ensuremath{\tau^+\to\pi^+\nu}}
\newcommand{\taurhox}{\ensuremath{\tau^+\to\rho^+\nu}}

\newcommand{\bptaunu}{\ensuremath{\Bp \to\taup \nu\ }}
\newcommand{\taurho}{\ensuremath{\tau^+\to\rho^+\nu\ }}
\newcommand{\taue}{\ensuremath{\tau^+\to e^+\nu\bar{\nu}\ }}
\newcommand{\taumu}{\ensuremath{\tau^+\to \mu^+\nu\bar{\nu}\ }}
\newcommand{\taupi}{\ensuremath{\tau^+\to\pi^+\nu\ }}

\def\mES       {\ensuremath{m_{\mbox{\scriptsize{ES}}}}}

\begin{document}

\begin{flushleft}
\babar-PUB-\BABARPubYear/\BABARPubNumber\\
SLAC-PUB-\SLACPubNumber\\
\end{flushleft}

\title{Evidence of \boldmath{\btn} decays with hadronic \textit{B} tags}
%
\author{J.~P.~Lees}
\author{V.~Poireau}
\author{V.~Tisserand}
\affiliation{Laboratoire d'Annecy-le-Vieux de Physique des Particules (LAPP), Universit\'e de Savoie, CNRS/IN2P3,  F-74941 Annecy-Le-Vieux, France}
\author{J.~Garra~Tico}
\author{E.~Grauges}
\affiliation{Universitat de Barcelona, Facultat de Fisica, Departament ECM, E-08028 Barcelona, Spain }
\author{A.~Palano$^{ab}$ }
\affiliation{INFN Sezione di Bari$^{a}$; Dipartimento di Fisica, Universit\`a di Bari$^{b}$, I-70126 Bari, Italy }
\author{G.~Eigen}
\author{B.~Stugu}
\affiliation{University of Bergen, Institute of Physics, N-5007 Bergen, Norway }
\author{D.~N.~Brown}
\author{L.~T.~Kerth}
\author{Yu.~G.~Kolomensky}
\author{G.~Lynch}
\affiliation{Lawrence Berkeley National Laboratory and University of California, Berkeley, California 94720, USA }
% ...
\collaboration{The \babar\ Collaboration}
\noaffiliation


\date{July 3, 2012}

\begin{abstract}
\noindent We present a search for the decay \btn\ using $467.8 \times 10^6$ \BB pairs   
collected at the \FourS resonance with the \babar\ detector 
at the SLAC PEP-II $B$-Factory. 
We select a sample of events with one completely reconstructed \Bm in the hadronic
decay mode ($B^-\to D^{(*)0}X^-$ and $\Bm \to \jpsi X^-$).
We examine the rest of the event to search for a \bptaunu decay. 
We identify the \taup lepton  in the following modes: $\tautoenunu$, $\tautomununu$,
$\tautopinu$ and $\tau^+ \to \rho^+ \overline{\nu}$.
We find an excess of events with respect to the expected background, which
excludes the null signal hypothesis at the level of 3.8$\sigma$ 
(including systematic uncertainties) and
corresponds to a branching fraction 
central value of  $\mathcal{B}(\btn)=( 1.83^{+0.53}_{-0.49}(\mbox{stat.}) \pm 0.24 (\mbox{syst.})) \times 10^{-4}$. 
\noindent\end{abstract}

\maketitle

The study of the purely leptonic decay \btn\ \cite{conj} is 
of particular interest to test the predictions of the Standard Model (SM) and 
to probe of new physics effects.
It is sensitive to the product of the $B$ meson decay constant $f_{B}$, 
and the absolute value of the Cabibbo-Kobayashi-Maskawa matrix element $\Vub$~\cite{ckm}.
In the SM the branching fraction is given by:
\begin{equation}
\label{eqn:br}
\mathcal{B}(B^{+} \rightarrow {\taup} \nu)= 
 \frac{G_{F}^{2} m^{}_{B}  m_{\tau}^{2}}{8\pi}
\left[1 - \frac{m_{\tau}^{2}}{m_{B}^{2}}\right]^{2} 
f_{B}^{2} \Vub^{2}
\tau_{\Bu}, 
\end{equation}
where  
$G_F$ is the Fermi constant,
$m^{}_{B}$ and $m_{\tau}$ are, the $\Bu$ meson and $\tau$ lepton masses, respectively,
and $\tau_{\Bu}$ is the $\Bu$ lifetime.
Using the Lattice QCD calculation of $f_B = (189 \pm 4)~\mev$~\cite{lattice}, and  the \babar\ measurement of \Vub\ from charmless semileptonic B exclusive 
decays~\cite{vubmeasexcl}, the predicted SM  value of the brancing fraction is $\mathcal{B}_{SM}(\btn) = (0.62 \pm 0.12) \times 10^{-4}$. If we use the \babar\ measurement of \Vub\ from inclusive charmless semileptonic B decays~\cite{vubmeasincl}, the SM prediction is 
$\mathcal{B}_{SM}(\btn) = (1.18 \pm 0.16) \times 10^{-4}$.

The process is sensitive to possible extensions of the SM. 
For instance, in two-Higgs doublet models~(2HDM)~\cite{twohiggs} and in minimal supersymmetric extensions~\cite{hightanbeta} it can be mediated by a charged Higgs boson.
A branching fraction measurement can, therefore, also be used to constrain the parameter space of new physics models.

The data used in this analysis were collected with the \babar\ detector
at the \pep2\ storage ring. 
The sample corresponds to an integrated
luminosity of \onlumi at the \FourS\ resonance.
The sample contains $(467.8 \pm 5.1) \times 10^{6}$ \BB decays ($N_{B\overline{B}}$). 
The detector is described in detail elsewhere~\cite{nimbabar}.
Charged particle trajectories are measured in the tracking system
composed of a five-layer double-sided silicon vertex tracker and a 40-layer drift chamber,
operating in a  1.5~T solenoidal magnetic field.
A Cherenkov detector is used for charged $\pi$--$K$ discrimination, a CsI calorimeter
for photon and electron identification, and 
the flux return of the solenoid, which consists of layers of iron 
interspersed with resistive plate chambers or limited streamer tubes, for muon
and neutral hadron identification.

We use a Monte Carlo (MC) simulation based on \geantfour~\cite{geant4}
to estimate signal selection efficiencies and to study backgrounds.
In MC simulated signal events, one \Bp meson decays as \btn  and the other decays in any final state.
The \BB\ and continuum MC samples are equivalent to approximately 3 times and
1.5 times the data sample, respectively.
Beam-related background and detector noise are sampled from data 
and overlaid on the simulated events.

We reconstruct an exclusive decay of one of the \B mesons in the event
(which we refer to as the tag-\B) and examine the rest of the event
for the experimental signature of \btn. The tag-\B reconstruction can be performed
by looking at both hadronic \B decays and semileptonic \B decays. 
Published results from both \babar\ and Belle are summarized in Table~\ref{pubres}.
\begin{table}[bt]
\begin{center}
\caption{Published results for \btn\ from \babar\ and Belle collaborations.}
\label{pubres}
\renewcommand{\arraystretch}{1.2}
\begin{tabular}{llc}
\hline
\hline
Experiment & Tag  & Branching Fraction ($\times 10^{-4})$ \\
\hline
\noalign{\vskip1pt}
\babar\  &  hadronic~\cite{babarhad0}       & $1.8 ^{+0.9}_{-0.8} \pm 0.4 \pm 0.2 $ \\
\babar\  &  semileptonic~\cite{babarsl}     & $1.7 \pm 0.8 \pm 0.2$ \\
Belle    &  hadronic~\cite{bellehad}        & $1.79 ^{+ 0.56}_{ -0.49}{}^{+ 0.46}_{ -0.51} $ \\
Belle    &  semileptonic~\cite{bellesl}     & $1.54 ^{+ 0.38}_{ -0.37}{}^{+ 0.29}_{ -0.31} $ \\
\hline
\hline
\end{tabular}
\end{center}
\end{table}

We reconstruct the tag-\B candidate in the set of hadronic decays $B^-\to M^0 X^-$, where $M^0$ denotes a $D^{(*)0}$ or a  \jpsi, and 
$X^-$ denotes a system of hadrons with total charge $-1$ composed of
$n_1 \pi^{\pm}$, $n_2 K^{\pm}$, $n_3 \pi^0$, $n_4 \KS$ where
$n_1 + n_2 \le 5$, $n_2$, $n_3$~and~$n_4 \le 2$.
We reconstruct the $D^0$ as $D^0  \to K^- \pi^+$, 
$K^- \pi^+ \pi^0$, $K^- \pi^+ \pi^- \pi^+$, $\KS \pi^0$, $\KS \pi^+ \pi^-$,  $\KS \pi^+ \pi^- \pi^0$, $K^+ K^-$, or~$\pi^+ \pi^-$.  
We reconstruct the \Dstarz meson as $\Dstarz \to D^0 \pi^0, D^0 \gamma$, and the \jpsi meson via their decays
$\jpsi \to e^+e^-, \mu^+\mu^-$.
Two kinematic variables are used to discriminate between correctly reconstructed tag-\B candidates 
and mis-reconstructed events:
the beam energy-substituted mass $\mes \equiv \sqrt{s/4 - p_B^2}$, and the
energy difference $\DeltaE \equiv E_B-\sqrt{s}/2$, where $\sqrt{s}$ is the
total energy in the $\Upsilon(4S)$ center-of-mass system (CM) and $p_B$ and
$E_B$ respectively denote the momentum and the energy of the tag-\B
candidate in the CM. The resolution on \DeltaE is
measured to be $\sigma_{\DeltaE} = 10-35 \mev$, depending on the decay
mode; we require $|\DeltaE| < 3\sigma_{\DeltaE}$. 
Events with a tag-\B candidate arise from two possible classes with different \mes distributions.
One class includes signal events with a correctly reconstructed tag-\B, and
background events from $\FourS \to \BpBm $ with a correctly reconstructed tag-\B.
All these events are  characterized by an \mes distribution peaked
at the nominal \B mass(signal and peaking background).
The other classes of events consist of continuum background, $\epem \to \q \qbar $ (\q = \u, \d, \s, \c) and $\epem \to \tautau$, 
and combinatorial background, $\Y4S \to \BzBzb$ or \BpBm in which the tag-\B is misreconstructed.
These events are characterized by a smooth \mes distribution.
 
\begin{figure}[!tbh]
\begin{center}
\begin{tabular}{c}
\includegraphics[width=\linewidth]{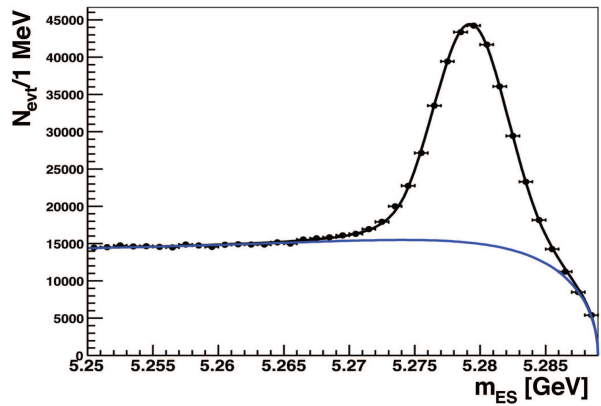}
\end{tabular}
\caption{Fit to the \mes distribution in data. Dots are data, the upper curve is the global fit result and the lower curve represents the fitted combinatorial and continuum background.}
\label{fig:mes_data}
\end{center}
\end{figure}

If multiple tag-\B candidates are reconstructed in the event, we select the one with the lowest value of $|\Delta E|$.
After the reconstruction of the tag-\B, we require the presence of only one well-reconstructed track (signal track),
with charge opposite to that of the tag-\B.
The purity ${\cal P}$ of each reconstructed tag-\B decay mode is estimated as the ratio of the number of peaking events with $m_{ES} > 5.27 \gev$ to the total number of events in the same range. 
The yield in data is determined by means of an extended unbinned maximum likelihood fit to the \mes distribution, as shown in Fig.~\ref{fig:mes_data}.
We use a phenomenologically motivated threshold function (ARGUS function~\cite{argus}) as probability density function (PDF) to describe the continuum and 
combinatorial background components in the fit,
while for the correctly reconstructed tag-\B component we use
a Gaussian distribution plus an exponential tail for the PDF (Crystal Ball function)~\cite{cball}.
We use only events with the tag-\B reconstructed in decay modes with ${\cal P} > 0.1$.
Combinatorial and continuum background distributions in any discriminating variable are estimated 
from a sideband in \mes ($ 5.209 \gev < \mes < 5.260 \gev$) and are extrapolated into the signal region ($\mes > 5.270 \gev$) using the
results of a fit to an ARGUS function.
The peaking \BpBm background shape is determined from \BpBm MC, after subtraction of
the combinatorial component to avoid double counting. 

The signal-side $\tau$ lepton is reconstructed in four decay modes: \tauex, \taumux, \taupix, and \taurhox, 
totaling approximately 70\% of all $\tau$ decays.
We separate the event sample into four categories using particle identification 
criteria applied to the signal track ($e^+$, $\mu^+$,  and $\pi^+$). 
The \taurho sample is obtained by associating the signal
track $\pi^+$ with a $\pi^0$ reconstructed from a pair of neutral clusters with an invariant mass
between 115 \mevcc and 155 \mevcc.   

In order to remove the $\epem \to \tautau$ background, we impose \mtau mode dependent
requirements on the ratio between the $2^{nd}$ and the $0^{th}$ Fox-Wolfram moments R2~\cite{r2}
calculated using all the tracks and neutral clusters of the event. This preserves 90\% of the \btn signal.

To reject continuum background, we use the absolute value of
$\cos \theta_{TB}$, the cosine of the angle in the CM frame between
the thrust axis~\cite{thrust} of the tag-\B and the thrust axis
of the remaining charged and neutral candidates in the event. 
For correctly reconstructed tag-B candidates the $\left|\cos \theta_{TB}\right|$
distribution is expected to be uniform,
while for jet-like $\epem \to \q \qbar $ continuum events it peaks strongly at 1.
In order to reject background from events with a correctly reconstructed tag-\B,
we study the distribution of several discriminating variables exploiting the 
different kinematics between the signal and background of the remaining reconstructed
candidates. 
We use the missing momentum polar angle in the laboratory frame 
$\vec{p}_{miss} = \vec {p}_{CM} - \vec{p}_{tag\B} - \vec{p}_{trk} - \sum_{neut} \vec{p}_i$,
where $\vec {p}_{CM}$ is the total momentum of the beams, $\vec{p}_{tag\B}$ is the 
reconstructed momentum of the tag-\B, and $\vec{p}_{trk}$ is the reconstructed track momentum,
and the sum is extended on all the neutral candidates reconstructed in the calorimeter
not assigned to the tag-\B.
For the \taupi mode, we combine 
$p^*_{trk}$ (where the star denotes the CM frame) and the cosine of the angle between $\vec{p}_{miss}$ and the beam axis ($\cos \theta_{miss}$) 
in a likelihood ratio 
\begin{equation}
\label{eq:llratio}
L_P =  \frac{L_S( p^*_{trk}, \cos \theta_{miss} )}{(L_S( p^*_{trk}, \cos \theta_{miss} ) +  L_B( p^*_{trk}, \cos \theta_{miss} ) )},
\end{equation}
where the signal (S) and background (B) likelihoods have been obtained from the product
of the PDFs of the two discriminating variables:
$L_S( p^*_{trk}, \cos \theta_{miss} ) = P_{S}( p^*_{trk}) P_{S}( \cos \theta_{miss} )$
and
$L_B( p^*_{trk}, \cos \theta_{miss} ) = P_{B}( p^*_{trk}) P_{B}( \cos \theta_{miss} )$.
Similarly, for the \taurho mode we combine  four discriminating
variables in the likelihood ratio $L_P$:  $\cos \theta_{miss}$,
the invariant mass of the \piz candidate, the $\rho^+$ candidate momentum, 
and the invariant mass of the 
$\pi^+ \piz$ pair used to make the $\rho^+$ candidate.
The PDFs used in the likelihood ratio for the signal and background are determined from signal and \BpBm MC samples, respectively.

The most powerful discriminating variable is \eextra, defined as the sum of the energies of
the neutral clusters not associated with the tag-\B or with the signal $\pi^0$
from the \taurho mode, and passing a minimum energy requirement (60 \mev).
Signal events tend to peak at low \eextra. Background events, 
which contain additional sources of neutral clusters, tend to be distributed at higher values.
The signal region in data is kept blind until the end of the analysis chain when we extract the signal yield, 
meaning that we do not use events in data with $\eextra < 400 \mev$ during the selection optimization procedure
and for the evaluation of background shapes.

We optimize the selection requirements, including those on the purity ${\cal P}$ of the tag-\B and
the minimum energy of the neutral clusters, minimizing the expected uncertainty in 
the branching fraction fit. In order to estimate the uncertainty, which includes
the statistical and the dominant systematic sources, we run 1000 MC simulated pseudo experiments
extracted from the background and signal expected \eextra distributions  for a set of possible selection requirements,
assuming a signal branching fraction of $1.8 \times 10^{-4}$~\cite{babarhad0}.
\begin{figure}[!tbh]
\begin{center}
\begin{tabular}{cc}
\multicolumn{2}{c}{
\includegraphics[width=\linewidth]{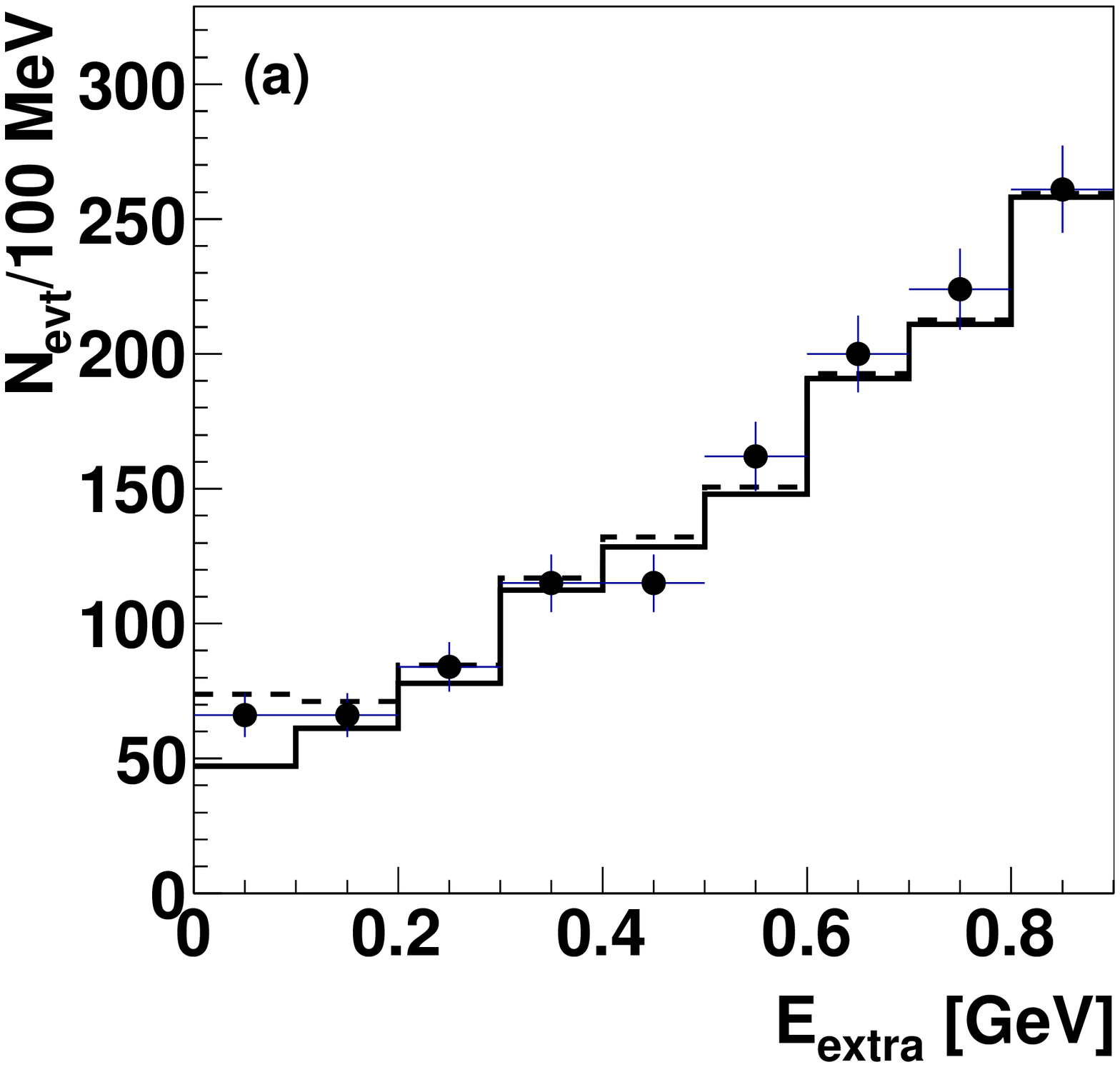}
} \\
\includegraphics[width=0.5\linewidth]{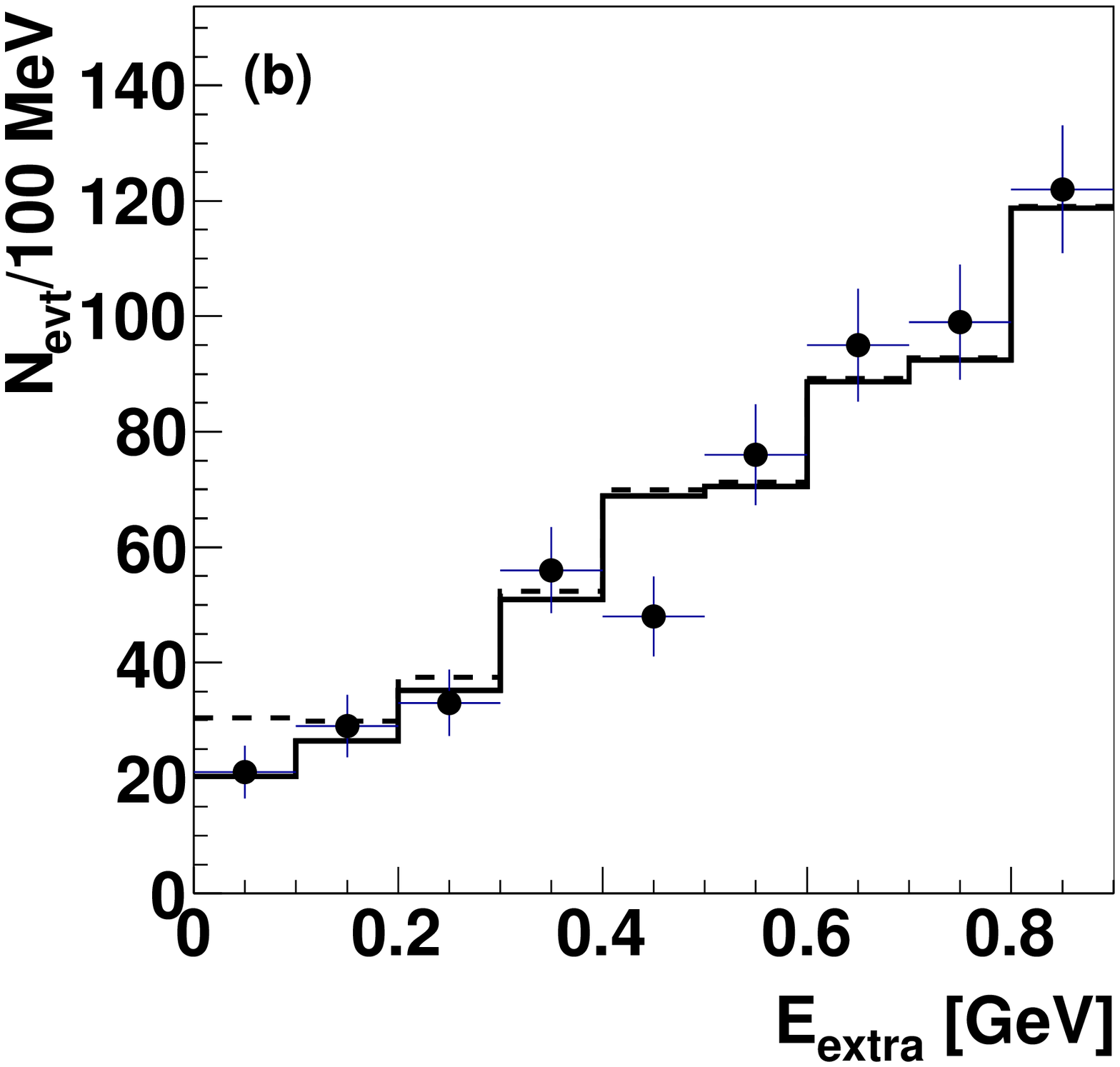} &
\includegraphics[width=0.5\linewidth]{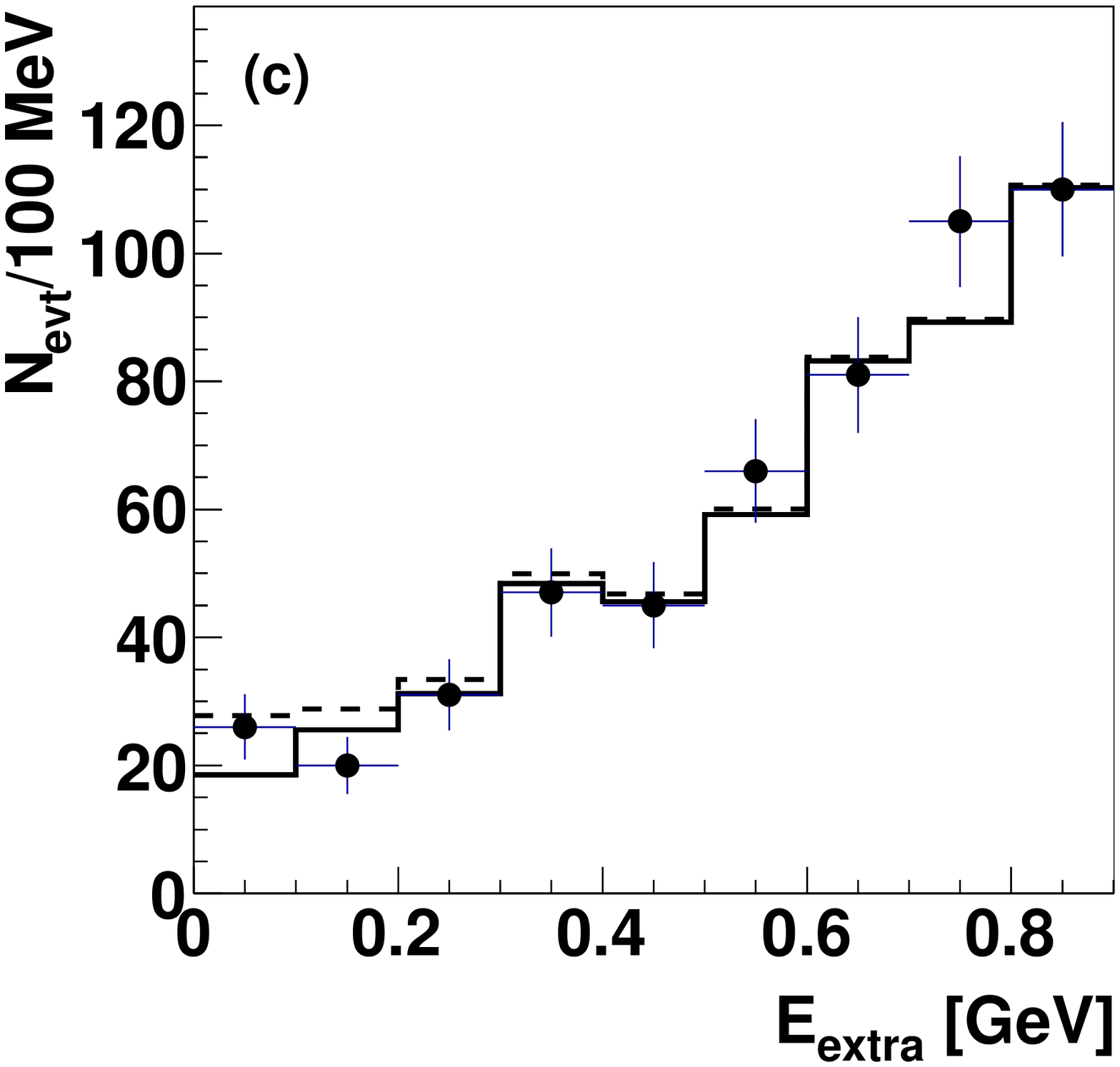} \\
\includegraphics[width=0.5\linewidth]{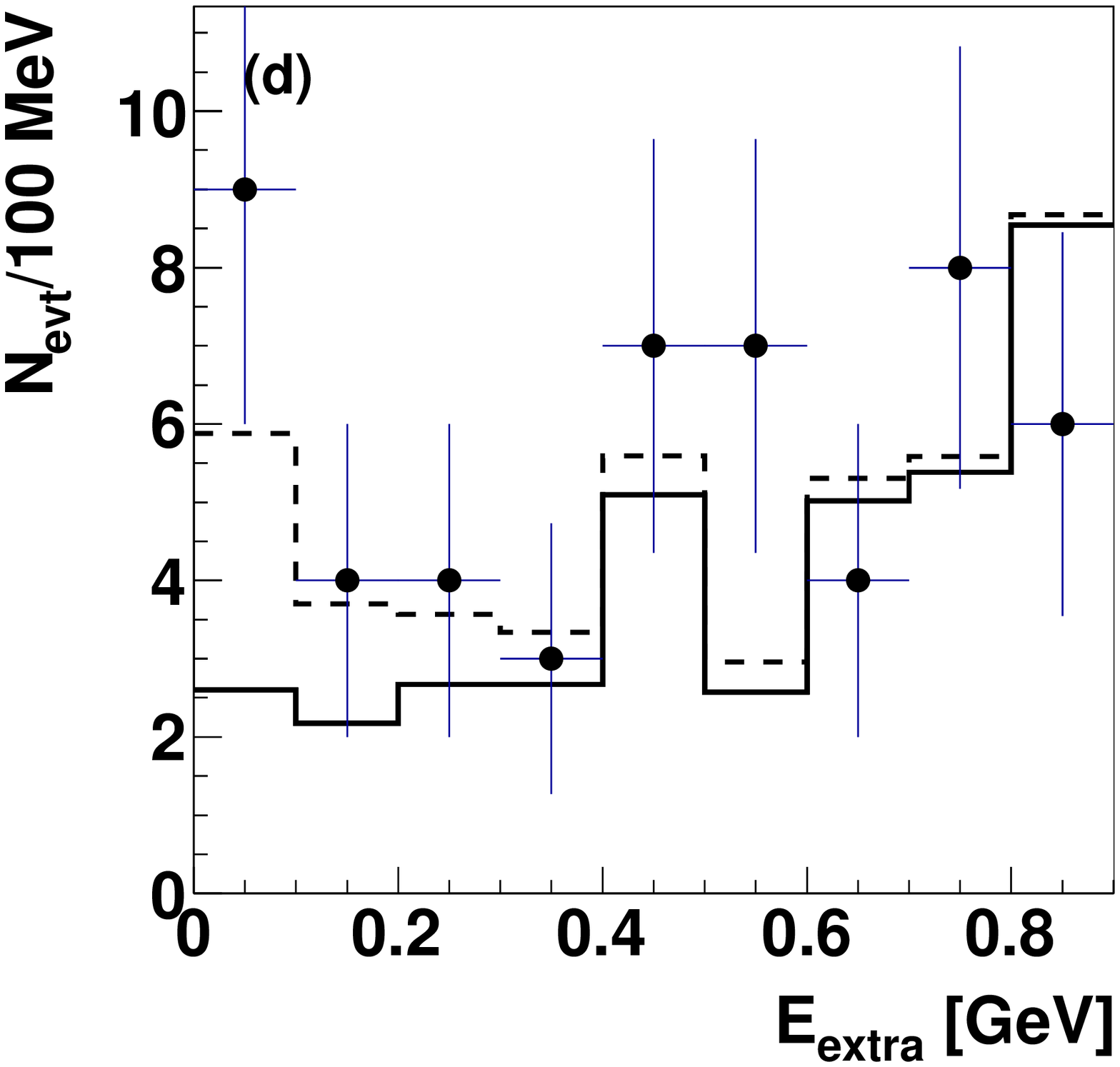} &
\includegraphics[width=0.5\linewidth]{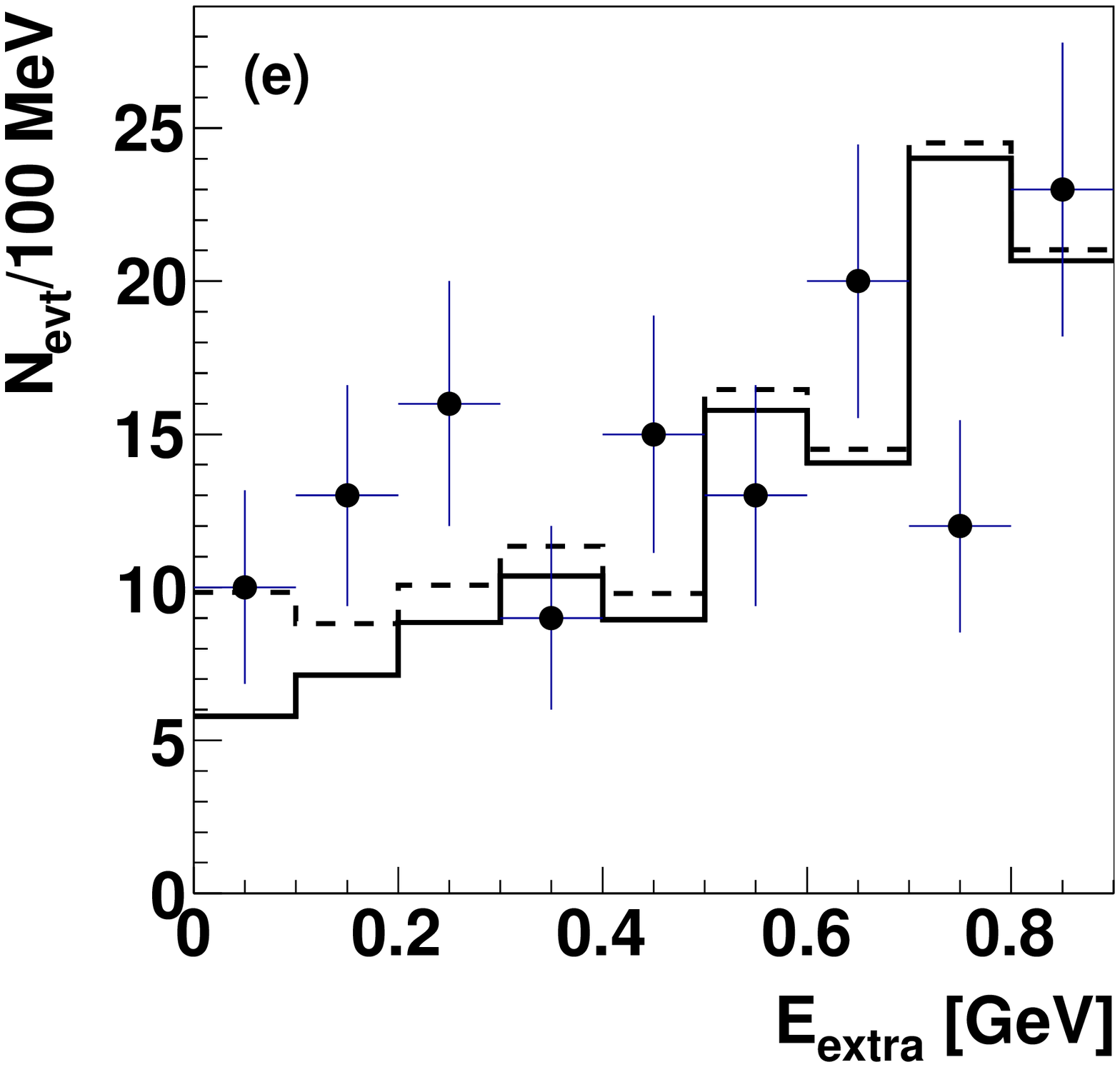} 
\end{tabular}
\caption{\eextra distribution in data (points with error bars) with all selection
 requirements applied and fit results overlaid. The hatched histogram is
 the background and the dashed component is the best-fit signal excess distribution. 
Plot (a) shows all $\tau$ decay modes fitted simultaneously. Lower plots show the projection of the
simultaneous fit result on the four analyzed $\tau$ decay modes: (b) \taue, \emph{(c)} \taumu,
(d) \taupi, (e) \taurho.}
\label{fig:eextrasel}
\end{center}
\end{figure}

Table \ref{tab:selectiontable} summarizes the signal selection requirements  
and Fig.~\ref{fig:eextrasel} shows the \eextra distribution with all the selection requirements applied.
The background events populating the low \eextra region are mostly semileptonic \B decays for the leptonic 
modes. For the \taupi mode the background is composed mostly of charmless hadronic \B decays and semileptonic \B decays with a muon in the final state. For the \taurho mode the backgrounds are charmed hadronic \B decays, semileptonic \B decays with a muon in the final state and a small fraction with a $\tau$.

\begin{table}[!tbh]
\begin{center}
\caption{Optimized signal selection criteria for each \mtau mode.}
\begin{tabular}{lcccc}
\hline
\hline 
 Variable &$e^+$&$\mu^+$&$\pi^+$&$\rho^+$\\
\hline
$\mathcal P$& \multicolumn{4}{c}{$>10\%$} \\ 
Cluster energy (\mev)&  \multicolumn{4}{c}{$>60$}  \\
$\mathcal{R}2$  &$<0.57$&$<0.56$&$<0.56$&$<0.51$\\
$|\cos \theta_{TB}|$ &$<0.95$ &  $<0.90$& $<0.65$&$<0.8$\\
$L_P$ &  &  & $>0.30$ &$>0.45$\\
\hline
\hline
\end{tabular}
\label{tab:selectiontable}
\end{center}
\end{table}

%
%
%
%
%
%
We use an extended unbinned maximum likelihood fit 
to the measured \eextra distribution to extract
the $B^+\to\tau^+\nu$ branching fraction.  
The likelihood function for the $N_k$ candidates  reconstructed in one of the four $\tau$ decay modes $k$ is
\begin{equation}
\label{eq:pdfsum}
{\cal L}_k = \frac{e^{-(n_{s,k}+n_{b,k}) }}{N_k !}\!\prod_{i=1}^{N_k}
                \bigg\{ n_{s,k} 
         {\cal P}_{k}^{s}(E_{i,k})
        + n_{b,k} {\cal P}_{k}^{b}(E_{i,k}) 
        \bigg\},
\end{equation}
where $n_{s,k}$ is the signal yield, $n_{b,k}$ is
the background yield,
$E_{i,k}$ is the \eextra value of the $i^{th}$ event, ${\cal P}_k^s$ is the
PDF of signal events, and ${\cal P}_k^b$ is the
PDF of background events.
The background yields in each decay mode are permitted to float independently of each other in the fit, while the 
signal yields are constrained to a single branching ratio via the relation:
\begin{equation}
\label{eq:BFcalc}
 n_{s,k} = N_{ B\overline{B}} \times \epsilon_k\times {\cal B}
\end{equation}
where $\epsilon_k$ is the reconstruction efficiency of a particular $\tau$ decay mode, and
${\cal B}$ is the $B^+\to\tau^+\nu$ branching fraction. 
The parameters $N_{ B\overline{B}}$ and $\epsilon_k$  are fixed in the
fit while ${\cal B}$ is allowed to vary.
The reconstruction efficiencies $\epsilon_k$, which include the $\tau$ branching fractions, are obtained from MC-simulated signal events (see Table~\ref{tab:res}).
Since the tag-\B reconstruction efficiency is included in $\epsilon_k$ and is estimated from the signal MC, we apply a correction factor
of $R_{\rm{data/MC}} = 0.926 \pm 0.010$
to take into account data/MC differences. This is derived from the
ratio of the peaking component of the \mES\ distribution for the hadronic tag-\B in data and in MC simulated events.

\begin{figure}[!tbh]
\begin{center}
\includegraphics[width=\linewidth]{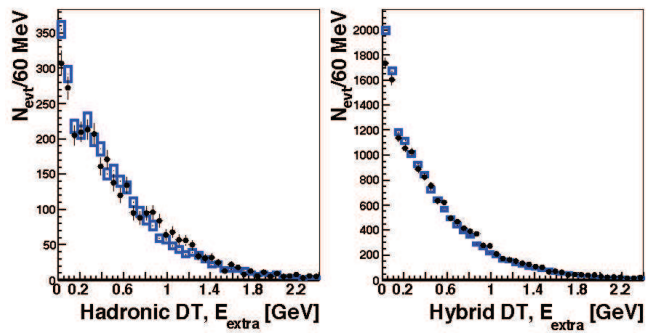}
\caption{\eextra distribution for double tagged events. The ``signal" \B is reconstructed in hadronic decays (left plot) or semileptonic decays (right plot). Points are data and boxes are MC simulation.}
\label{fig:doubletags}
\end{center}
\end{figure}

The signal PDF is obtained from a high statistics signal sample of MC
simulated data. 
We use a sample of fully reconstructed events to correct the signal PDF for data/MC disagreement
In addition to the reconstructed tag-\B, a second
\B is reconstructed in the hadronic or the semileptonic decay mode using
tracks and neutral clusters not assigned to the tag-\B.
In order to estimate the correction to the signal PDF, we compare the
distribution of \eextra in this double tagged event sample from
experimental data and MC simulations. 
The MC distributions are normalized to the experimental data and the comparison is shown in Fig.~\ref{fig:doubletags}.
We extract the correction function by taking the ratio of the two
distributions and fitting it with a second order polynomial.

We determine the PDF of the combinatorial background from the \mes sideband. 
The normalization of this component  in the signal region is  obtained by fitting the \mes~distribution after the selection has been applied.
The shape of the peaking background is taken from \BpBm MC.
The two background components  are added together into a single background PDF. 
We estimate the branching fraction by minimizing $-\ln {\cal L}$, where
${\cal L}  = \Pi_{k=1}^{4} {\cal L}_k$, and
${\cal L}_k$ is given in Eq.~\ref{eq:pdfsum}. The projections of the fit results are shown in Fig.~\ref{fig:eextrasel}.

We observe an excess of events with respect to the expected
background level and measure a branching fraction of 
${\cal B}( \btn) = ( 1.83 ^{+0.53}_{-0.49} )\times 10^{-4}$, where the
uncertainty is statistical. Table \ref{tab:res} summarizes the results from the fit.
We evaluate the significance of the observed signal, including only
statistical uncertainty, as $S = \sqrt{2 \ln({\cal L}_{s+b}/{\cal L}_{b})}$, where
${\cal L}_{s+b}$ and ${\cal L}_{b}$ denote the obtained maximum likelihood values in the signal
and background, and the background only hypotheses, respectively. 
We find $S = 4.2\sigma$.

\begin{table}
\caption{Reconstruction efficiency $\epsilon$, measured branching fractions, and statistical uncertainty
obtained from the fit with all the modes separately and constrained to
the same branching fraction. The $\tau$ decay mode branching fractions are included in the efficiencies.}
\label{tab:res}
\begin{center}
\begin{tabular}{lccc}
\hline
\hline
Decay Mode    &  $\epsilon_k (\times 10^{-4})$ & Signal yield & $\mathcal{B} (\times 10^{-4})$ \\
\hline
\noalign{\vskip1pt}
\taue  &   $2.47\pm0.14$     & $4.1 \pm 9.1$ & $0.35 ^{+0.84}_{-0.73}$ \\
\taumu &  $2.45\pm0.14$    & $12.9 \pm 9.7$  & $1.12 ^{+0.90}_{-0.78}$ \\
\taupi &  $0.98\pm0.14$      & $17.1 \pm 6.2$ & $3.69   ^{+1.42}_{-1.22}$ \\
\taurho & $1.35\pm0.11$     & $24.0 \pm 10.0 $ & $3.78   ^{+1.65}_{-1.45}$ \\
\noalign{\vskip1pt}
\hline
\noalign{\vskip1pt}
combined &  & $ 62.1 \pm 17.3$ & $1.83^{+0.53}_{-0.49}$ \\ 
\hline
\noalign{\vskip1pt}
\hline
\end{tabular}
\end{center}
\end{table} 

%
%
%
%
%
%
%

Additive systematic uncertainties are due to the uncertainties in the signal and background \eextra 
PDF shapes used in the fit.
To estimate the systematic uncertainty in the background PDF shape 
we repeat the fit of the branching fraction with 1000 variations of the background PDFs, 
varying each bin content within its statistical uncertainty.
We use the range of fitted branching fractions covering 68\% of the distribution
as systematic uncertainty
yielding an overall contribution of 10\%. 
We correct the systematic effects of disagreements between data and MC \eextra distributions
for signal events using a sample of completely reconstructed events in data and MC, as already 
described. 
To estimate the related systematic uncertainties, 
we vary the parameters of the second-order polynomial defining the correction within their uncertainty
and repeat the fit to the \btn branching fraction. We observe a 2.6\% 
variation that we take as the systematic uncertainty on the signal shape. 
Including the effects of additive systematic uncertainties, the significance of the result 
is evaluated as 3.8$\sigma$.

Multiplicative systematic uncertainties on the efficiency stem from 
the uncertainty in the tag-\B efficiency correction (5.0\%), 
electron identification (2.6\%),  muon identification (4.7\%), charged kaon veto (0.4\%),
and the finite signal MC statistics (0.8\%).
Table \ref{tab:systematics} summarizes
the systematic uncertainties. 
The total systematic uncertainty is obtained by combining all sources in quadrature.

\begin{table}
\caption{Contributions to systematic uncertainty on the branching
 fraction.}
\label{tab:systematics}
\begin{center}
\begin{tabular}{lc}
\hline
\hline
Source of systematics   & ${\cal B}$ uncertainty (\%)\\
\hline
Additive \\
\hline
Background PDF            &  10    \\
Signal PDF                     &  2.6   \\
\hline
Multiplicative \\
\hline
Tag-\B efficiency            &   5.0 \\  
\B counting                      &  1.1\\
Electron identification   &  2.6  \\  
Muon identification       &  4.7        \\  
Kaon identification        &   0.4\\
Tracking                        &   0.5  \\
MC statistics                 &  0.6        \\
\hline
Total                               &  13  \\
\hline
\hline
\end{tabular}
\end{center}
\end{table}

In summary, we have measured the branching fraction of the decay \btn
using a tagging algorithm based on the reconstruction of hadronic \B
decays using a data sample of $467.8 \times 10^6$ \BB pairs collected
with the \babar\ detector at the \pep2 $B$-Factory.
We measure the branching fraction to be 
$\mathcal{B}(\btn)=( 1.83^{+0.53}_{-0.49}(\mbox{stat.}) \pm 0.24 (\mbox{syst.})) \times 10^{-4}$,
excluding the null hypothesis by 3.8$\sigma$.
(including systematic uncertainty).
 This result supersedes our previous result using the same technique~\cite{babarhad0}.
Combining this result with the other \babar\ measurement of
$\mathcal{B}(\btn)$  derived from a statistically
independent sample~\cite{babarsl}, we obtain 
$\mathcal{B}(\btn)=( 1.79 \pm 0.48 ) \times 10^{-4}$, where
both statistical and systematic uncertainties are combined in quadrature.

\begin{figure}[!tbh]
\begin{center}
\begin{tabular}{cc}
\multicolumn{2}{c}{
\includegraphics[width=\linewidth]{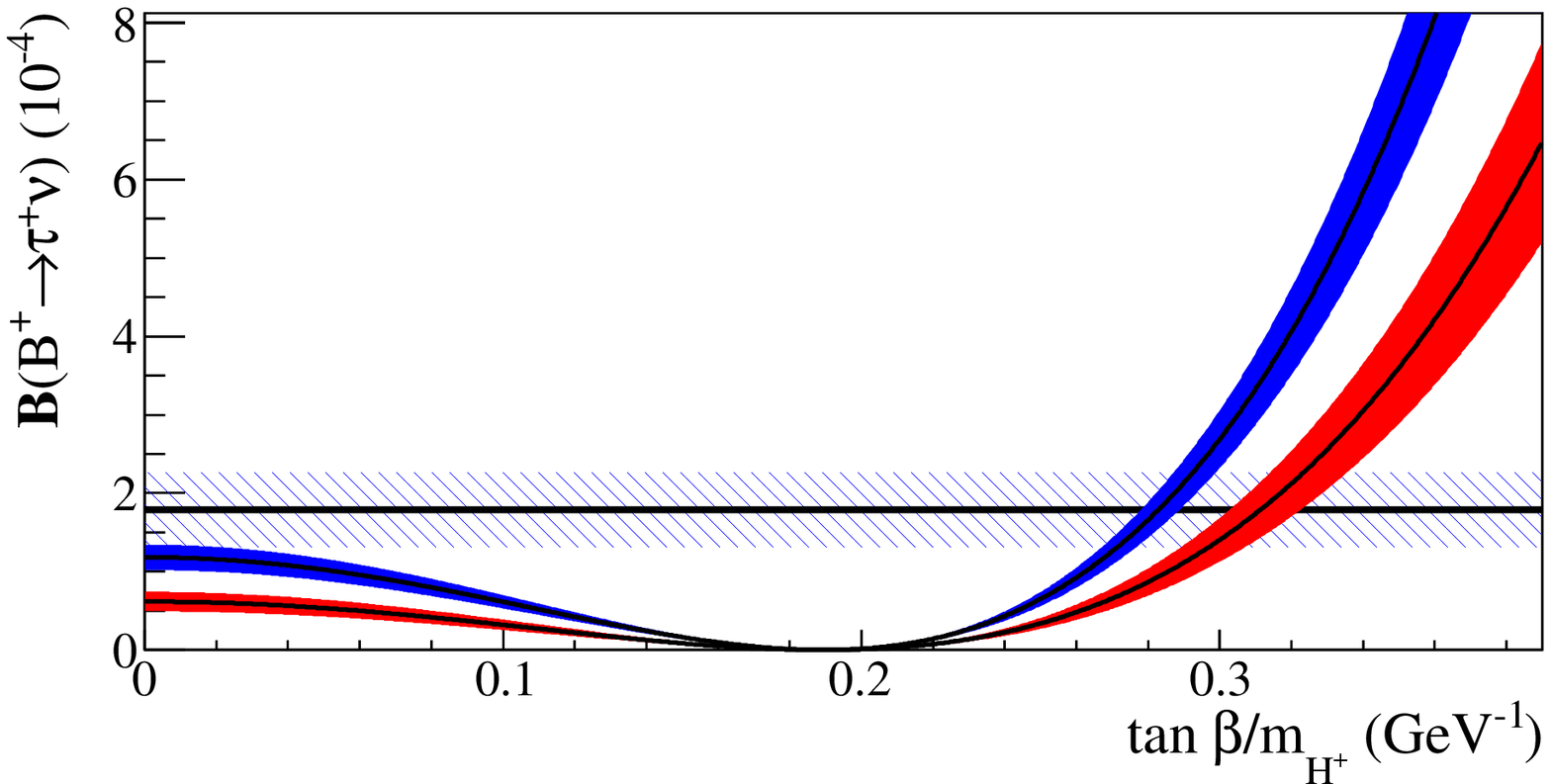}
} \\
\includegraphics[width=0.5\linewidth]{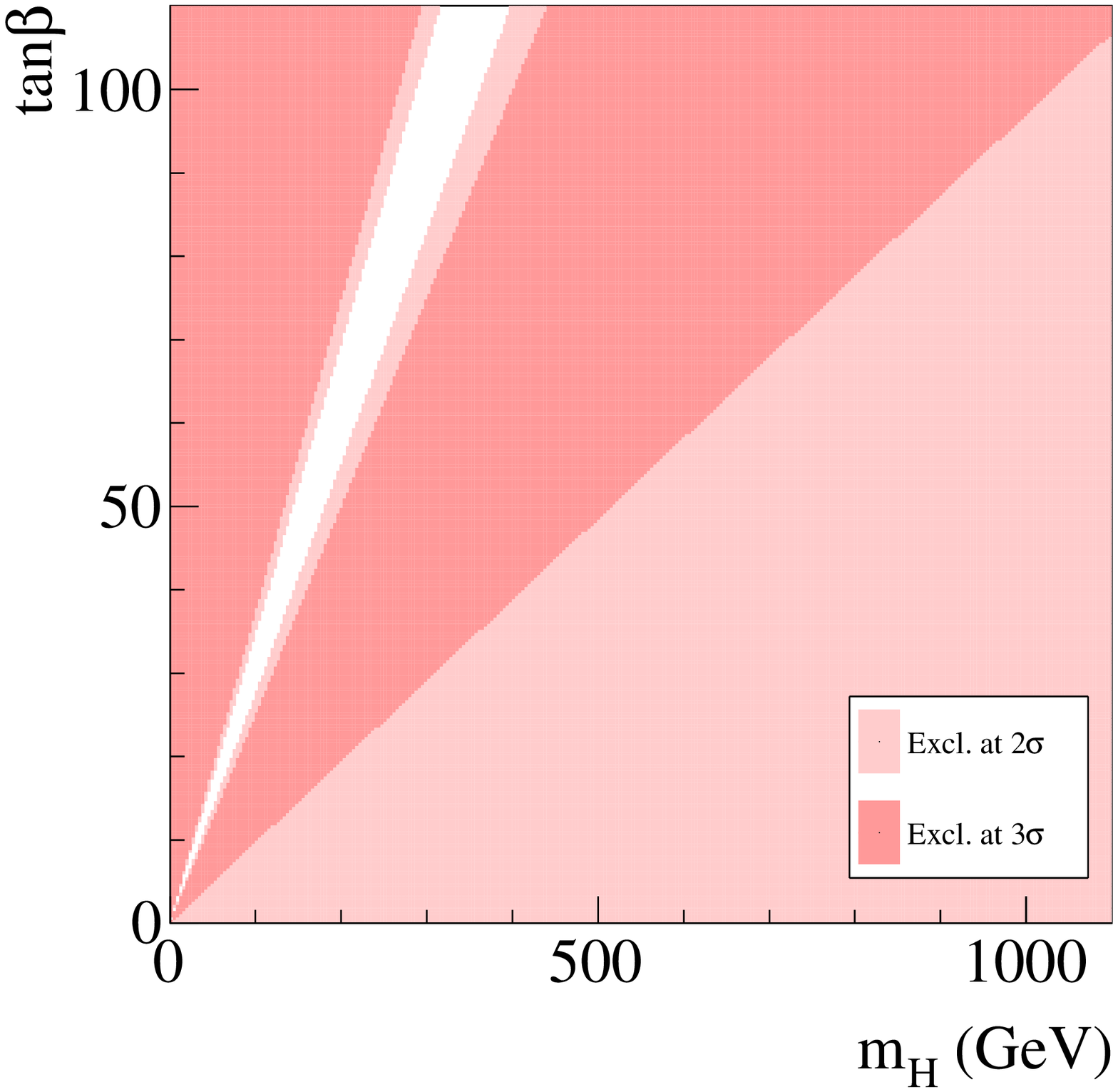} &
\includegraphics[width=0.5\linewidth]{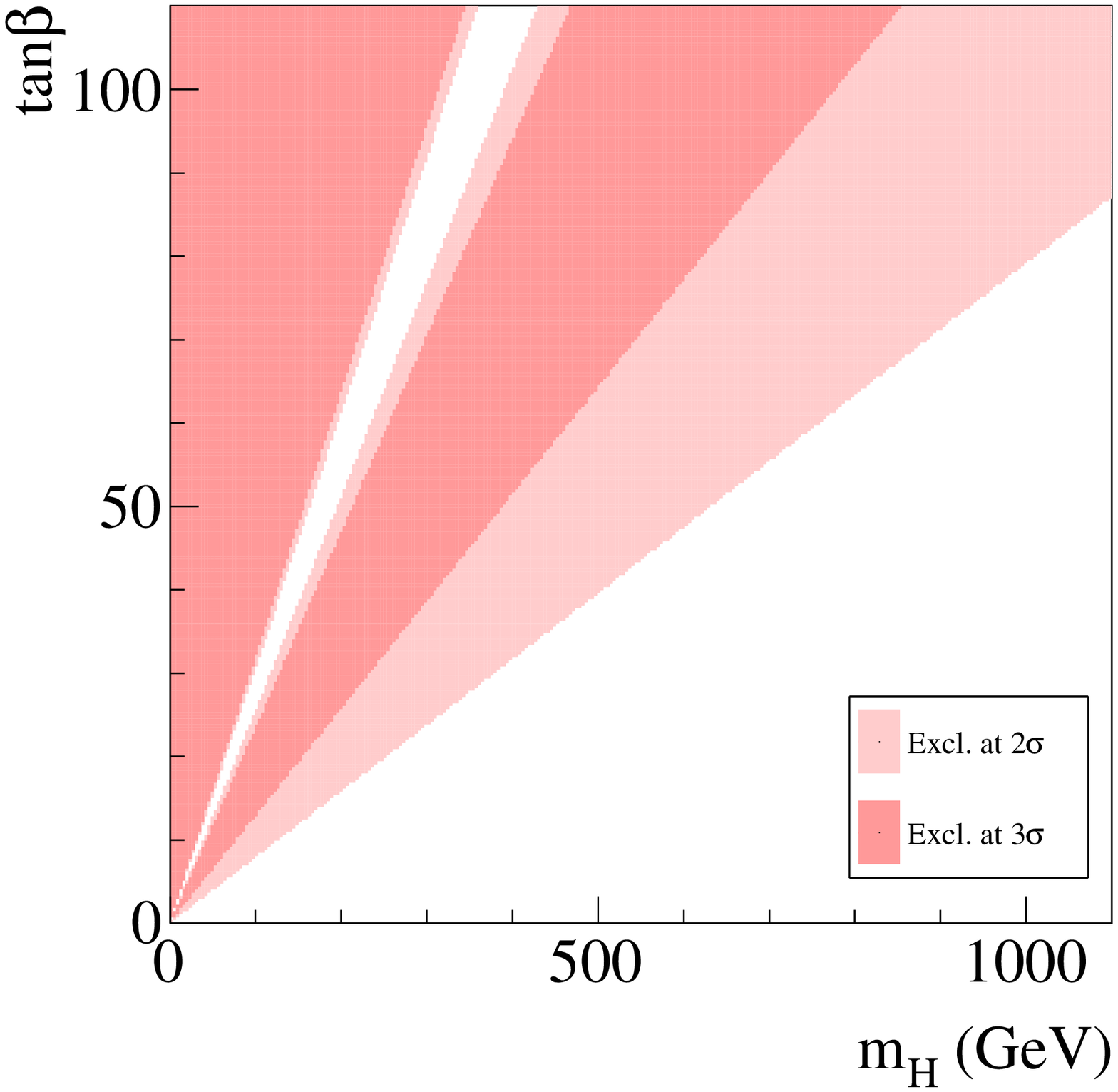} \\
\end{tabular}
\caption{Top plot: Comparison between the measured $\mathcal{B}(\btn)$ branching fraction (horizontal band) with the prediction of the 2HDM as a function of $\tan \beta / m_{H^+}$,
using exclusive (red/light gray) or inclusive (blue/dark gray) \Vub measurement.
Bottom plots: 90\% and 99\% C.L. exclusion regions in the $(m_{H^+},\tan \beta)$ plane using the exclusive (left) and inclusive (right) measurements of
\Vub.  }
\label{fig:twohiggsDM}
\end{center}
\end{figure} 

Our measurement of the branching fraction $\mathcal{B}(\btn)$ exceeds
the prediction of the SM determined using the values of \Vub\ extracted 
from exclusive semileptonic events and from inclusive
semileptonic events by $2.4 \sigma$ and $1.6 \sigma$, respectively.
We also determine,  separately for the exclusive and inclusive \Vub\ \babar\ measurements,
90\% C.L. exclusion regions in the parameter space 
of the 2HDM - type~II $(m_{H^+},\tan \beta)$, where $m_{H^+}$ is the charged Higgs
mass and $\tan \beta$ is the ratio of the vacuum expectation values of the two Higgs doublets.
We find that, taking \Vub\ from the exclusive measurement, 
most of the parameters space is excluded at 90\% C.L.
Using the higher value of \Vub\ from the inclusive measurement, 
the constraints are less stringent but already set a lower limits at the TeV 
scale for high $\tan \beta$. The same implications on 2HDM are supported by
a recent \babar\ study of the $\mathcal{B}(B \rightarrow D^{(*)} \tau \nu)$ decays \cite{btodtaunu}.
Fig.~\ref{fig:twohiggsDM} shows a comparison between the measured $\mathcal{B}(\btn)$ branching fraction with the prediction of the 2HDM as a function of $\tan \beta/m_{H^+}$ and the exclusion plots in the $(m_{H^+},\tan \beta)$ plane for the
exclusive and inclusive measurements of \Vub\ .

We are grateful for the 
extraordinary contributions of our \pep2\ colleagues in
achieving the excellent luminosity and machine conditions
that have made this work possible.
The success of this project also relies critically on the 
expertise and dedication of the computing organizations that 
support \babar.
The collaborating institutions wish to thank 
SLAC for its support and the kind hospitality extended to them. 
This work is supported by the
US Department of Energy
and National Science Foundation, the
Natural Sciences and Engineering Research Council (Canada),
the Commissariat \`a l'Energie Atomique and
Institut National de Physique Nucl\'eaire et de Physique des Particules
(France), the
Bundesministerium f\"ur Bildung und Forschung and
Deutsche Forschungsgemeinschaft
(Germany), the
Istituto Nazionale di Fisica Nucleare (Italy),
the Foundation for Fundamental Research on Matter (The Netherlands),
the Research Council of Norway, the
Ministry of Education and Science of the Russian Federation, 
Ministerio de Ciencia e Innovaci\'on (Spain), and the
Science and Technology Facilities Council (United Kingdom).
Individuals have received support from 
the Marie-Curie IEF program (European Union), the A. P. Sloan Foundation (USA) 
and the Binational Science Foundation (USA-Israel).

\bibliographystyle{apsrev}

\begin{thebibliography}{99}
\bibitem{conj} Charge-conjugate modes are implied throughout the paper.
\end{thebibliography}

\end{document}